\documentclass[conference]{IEEEtran}
\IEEEoverridecommandlockouts
\usepackage{cite}
\usepackage{amsmath,amssymb,amsfonts}
\usepackage{algorithmic}
\usepackage{graphicx}
\usepackage{textcomp}
\usepackage{xcolor}
\usepackage{hyperref}
\usepackage{booktabs}
\usepackage{newtxtext,newtxmath}
\usepackage{pdfpages}

\usepackage{fancyhdr}
\newcommand{\revise}[1]{\textcolor{black}{#1}}
\def\BibTeX{{\rm B\kern-.05em{\sc i\kern-.025em b}\kern-.08em
    T\kern-.1667em\lower.7ex\hbox{E}\kern-.125emX}}
\begin{document}

\title{Two-Level vs. Multi-Level Modelling: An Empirical Study of Cascading Maintenance Burden
}

\author{\IEEEauthorblockN{1\textsuperscript{st} Yuhong Fu}
\IEEEauthorblockA{
\textit{Adelaide University}\\
Mawson Lakes, Australia \\
0000-0003-2093-2326}
\and
\IEEEauthorblockN{2\textsuperscript{nd} Weixing Zhang}
\IEEEauthorblockA{
\textit{Karlsruhe Institute of Technology}\\
Karlsruhe, Germany \\
0000-0003-2890-6034}
\and
\IEEEauthorblockN{3\textsuperscript{rd} Bowen Jiang}
\IEEEauthorblockA{
\textit{Karlsruhe Institute of Technology}\\
Karlsruhe, Germany \\
0009-0001-8168-0866}
\and
\IEEEauthorblockN{4\textsuperscript{th} Haowei Cheng}
\IEEEauthorblockA{
\textit{Waseda University}\\
Tokyo, Japan \\
0009-0008-2265-6437}
\and
\IEEEauthorblockN{5\textsuperscript{th} Georg Grossmann}
\IEEEauthorblockA{
\textit{Adelaide University}\\
Mawson Lakes, Australia \\
0000-0003-4415-2228}
\and
\IEEEauthorblockN{6\textsuperscript{th} Karamjit Kaur}
\IEEEauthorblockA{
\textit{Adelaide University}\\
Mawson Lakes, Australia \\
0000-0003-0255-1060}
\and
\IEEEauthorblockN{7\textsuperscript{th} Matt Selway}
\IEEEauthorblockA{
\textit{Adelaide University}\\
Mawson Lakes, Australia \\
0000-0001-6220-6352}
\and
\IEEEauthorblockN{8\textsuperscript{th} Markus Stumptner}
\IEEEauthorblockA{
\textit{Adelaide University}\\
Mawson Lakes, Australia \\
0000-0002-7125-3289}
}

\maketitle
\thispagestyle{fancy}

\begin{abstract}
When a core definition in a software system changes, every dependent artefact must be updated to remain consistent. This cascading update problem pervades software maintenance and recurs in Model-Driven Engineering (MDE), where the evolution of a metamodel forces every conforming model to be migrated in lockstep. The dominant MDE paradigm, two-level modelling (2LM), fragments domain knowledge across two artefacts that must be kept mutually consistent, making this co-evolution a source of inconsistencies and maintenance effort. An alternative paradigm, multi-level modelling (MLM), unifies these artefacts into one and is claimed in the literature to reduce co-evolution burden. \revise{Prior empirical work has examined MLM's applicability and modelling effort, but its effect on co-evolution and maintenance burden has not yet been tested in a controlled, paired comparison against 2LM.}


We hypothesise that, for semantically equivalent evolution scenarios, MLM's structural unification yields fewer post-change inconsistencies and a smaller modification footprint than the 2LM split-artefact arrangement. To test this hypothesis, we present a pre-registered, mutation-based empirical comparison of co-evolution behaviour in 2LM and MLM; empirical results will follow in Stage 2. From a curated corpus of published 2LM co-evolution scenarios, we construct semantically equivalent MLM counterparts, apply identical evolution mutations to both, and measure outcomes through automated consistency checking and pre-registered hypothesis tests. Positive controls and a blinded mapping protocol guard against bias favouring MLM. This design provides the first empirical framework for assessing whether paradigm-level structural choices affect the cascading maintenance burden in model-based systems, operationalising co-evolution burden as two automatically measurable outcome variables, and delivering a reusable benchmarking protocol to support replication and extension.

\end{abstract}

\begin{IEEEkeywords}
Software Maintenance, Cascading Update, Model Co-Evolution, Model-Driven Engineering, Controlled Experiment
\end{IEEEkeywords}

\section{Introduction}\label{sec:intro}

When a core definition in a software system changes, every artefact that depends on it must be updated to remain consistent. This cascading update problem is a defining concern of software maintenance, recognised since Lehman's laws of continuing change~\cite{1456074}: database schema evolution forces application-layer migration~\cite{DBLP:journals/infsof/Sjoberg93}, and API breaking changes propagate to client code, tests, and downstream libraries~\cite{DBLP:conf/sigsoft/RobbesLR12,DBLP:journals/smr/DigJ06}. In each case, much of the cost lies not in the original change but in the cascade of dependent updates it triggers.

The same cascading update recurs in Model-Driven Engineering (MDE)~\cite{DBLP:journals/computer/Schmidt06,DBLP:series/synthesis/2017Brambilla}, a software engineering subfield in which models serve as primary development artefacts from which code, configurations, and analyses are derived. In MDE, the core definition of a system resides in its \textit{metamodel}, while one or more conforming \textit{models} populate that vocabulary with concrete instances. When the metamodel evolves, every conforming model must be migrated in lockstep.
This \textit{metamodel--model co-evolution} is the MDE-specific instance of the broader cascading update problem, and it is a source of inconsistencies and maintenance effort~\cite{DBLP:conf/edoc/CicchettiREP08,DBLP:conf/ecoop/HerrmannsdoerferBJ09,DBLP:journals/jss/KhelladiBHG17}. 
MDE underpins long-lived industrial systems in domains such as automotive, avionics, and energy infrastructure, where models drive code generation, simulation, and certification, 
and co-evolution costs accumulate over decade-long maintenance cycles. 
The dominant MDE paradigm, \textit{two-level modelling} (2LM), embodies this metamodel-and-model split precisely. An alternative paradigm, \textit{multi-level modelling} (MLM), unifies the metamodel and the model into a single artefact that internally spans multiple ontological levels, and is claimed in the literature to thereby reduce the burden of co-evolution~\cite{DBLP:journals/tosem/LaraGC14, DBLP:journals/bise/Frank14, DBLP:journals/dke/SelwaySMJGS17, DBLP:conf/in4pl/FuGKSS22, DBLP:conf/models/FuGKSS23}. The technical machinery of both paradigms, including the SLICER realisation of MLM used in this study, is detailed in Section~\ref{sec:background}.

\revise{Prior work has begun to evaluate MLM empirically: its applicability and modelling complexity benefits have been assessed against 2LM in specific settings~\cite{DBLP:journals/tosem/LaraGC14}. However, to our knowledge no controlled study has isolated the \emph{co-evolution/maintenance} dimension by applying identical evolution changes to paired 2LM and MLM artefacts and measuring the resulting inconsistency and modification burden~\cite{zhang2024supporting}. Existing maintenance-oriented evidence remains largely conceptual or based on illustrative examples,} leaving practitioners without guidance on whether MLM's purported maintenance benefits materialise in practice.

We address this gap with a pre-registered, mutation-based empirical comparison of co-evolution behaviour in 2LM and MLM. Drawing on a curated corpus of published 2LM co-evolution scenarios, we construct semantically equivalent MLM counterparts under a pre-registered mapping protocol, subject both versions to identical evolution mutations, and measure post-change inconsistencies and modification effort using automated consistency checking. The use of pre-registration, paired statistical analysis, positive controls, and a blinded mapping protocol jointly safeguards the comparison against bias favouring MLM. This Stage~1 report contributes: (i) a pre-registered design for comparing the maintenance burden of two structurally distinct ways of organising an evolving system's artefacts; (ii) an operationalisation of co-evolution burden as two automatically measurable outcomes, post-change inconsistency count and restoration modification count, each at its paradigm's native edit granularity; and (iii) a reusable benchmarking protocol (corpus, mappings, mutation taxonomy, analysis pipeline) supporting replication and extension. 

\section{Background}
\label{sec:background}

This section provides the technical background needed to understand the design choices in the rest of the paper. 

\paragraph{\textbf{Two-Level Modelling}}
The dominant MDE paradigm is two-level modelling (2LM), embodied by frameworks such as the Eclipse Modelling Framework (EMF) and the OMG's Meta-Object Facility (MOF). In 2LM, a metamodel (e.g., an Ecore file) defines the domain vocabulary, 
while a conforming model populates this vocabulary with concrete instances. This two-level separation fragments domain knowledge across two artefacts that must remain mutually consistent: any change to the metamodel (e.g., renaming a class, adding a mandatory attribute) typically obliges a corresponding migration of every conforming model~\cite{DBLP:conf/ecoop/HerrmannsdoerferBJ09, DBLP:journals/jss/KhelladiBHG17}.

\paragraph{\textbf{Multi-Level Modelling and SLICER}}
Multi-level modelling (MLM) is an alternative MDE paradigm. 
Rather than splitting domain knowledge across a metamodel and a model, MLM fixes a single generic linguistic metamodel (a small, stable kernel defining object, attribute, and relationship) and represents all domain knowledge in one artefact spanning as many ontological levels as the domain requires. Within it, an intermediate-level object can simultaneously instantiate the level above and type the level below. For example, PumpModel, CentrifugalPump-X200 (which instantiates PumpModel yet itself acts as a type), and the physical Pump-Serial-7341 can coexist in one artefact under one linguistic metamodel.

In SLICER~\cite{DBLP:journals/dke/SelwaySMJGS17}, a relationship-based MLM framework, this dual role is realised through five explicit inter-object relations: \textbf{standard instantiation} (InstN), \textbf{instantiation with extension} (InstX), and \textbf{specialisation by extension} (SpecX) introduce new ontological levels, while \textbf{specialisation by refinement} (SpecR) and \textbf{subset-by-specification} (SbS) operate within a level.
The level structure of a model thereby emerges dynamically from the combined network of these relations. 
The reference implementation is in DoME, a Smalltalk-based environment providing built-in validation and structural diff over SLICER artefacts~\cite{fu2024modelling}.

We select SLICER on three methodological grounds: (i) it provides an explicit, published mapping procedure from 2LM artefacts~\cite[Section~6]{DBLP:journals/dke/SelwaySMJGS17}, a hard prerequisite for the blinded mapping protocol and round-trip check (C3, Section~\ref{sec:hypotheses}); (ii) DoME offers a scriptable interface to both validation and structural diff, enabling automated measurement of InconsCount and ModCount; (iii) its relation-based level mechanism minimises the gap between mapping protocol primitives and measured artefact primitives. \revise{We treat SLICER as a representative MLM realisation because it instantiates the three structural properties that define the paradigm: arbitrary ontological levels emerging from the level-introducing relations, cross-level propagation of intensional content along instantiation and specialisation chains, and the dual instantiation/classification role of intermediate-level objects. The confirmatory claim is accordingly scoped to SLICER-realised MLM. Whether the measured effects transfer to potency-based or formal-logic realisations such as LML~\cite{DBLP:conf/modellierung/AtkinsonG16}, FMMLx~\cite{DBLP:journals/bise/Frank14}, MultEcore~\cite{DBLP:conf/models/MaciasRS16}, or DeepTelos~\cite{DBLP:conf/er/JeusfeldN16} is an external-validity question we defer to the multi-framework replication outlined in Section~\ref{sec:future}.} Bias risks introduced by this selection are addressed by the Mapping Protocol Bias and Tool Maturity controls detailed in Section~\ref{sec:execution_plan}.

\paragraph{\textbf{Co-Evolution as Intra-Model Maintenance}}
MLM reframes co-evolution without eliminating it: an upper-level edit still leaves dependent lower-level objects inconsistent, but the inconsistency is now governed by the fixed linguistic metamodel and checkable within one tool rather than across two separately maintained files. Whether this reframing yields measurable reductions in inconsistency risk and modification effort is the question this study tests.

\section{Hypotheses}\label{sec:hypotheses}
We operationalise the conjectured advantage of MLM along two dimensions of co-evolution burden: the \textbf{risk of inconsistency} an evolution step introduces, and the \textbf{effort} needed to absorb it. We state one primary hypothesis per dimension. Both are one-sided, predicting that MLM's artefact unification reduces the measured quantity. The null hypotheses are tested at $\alpha = 0.05$ with multiple-comparison correction across the two tests (Section~\ref{sec:execution_plan}). Following the RR Author's Guide, each hypothesis carries a pre-specified positive control (C1, C2), and a shared mapping control (C3) covers both.

\paragraph{\textbf{H1 --- Inconsistency risk}}
\textbf{Hypothesis.} For semantically equivalent scenarios, identical evolution mutations yield, on average, fewer post-change inconsistencies under MLM than under 2LM, each count measured under its paradigm's native consistency checking.

\textbf{Rationale.} In 2LM, inter-artefact synchronisation can lag or remain incomplete, generating inconsistencies after an evolution step (Section~\ref{sec:background}). In MLM the same step occurs within one artefact under a fixed linguistic metamodel, so no separate artefact can fall out of sync and the conditions for inconsistency are narrower. We therefore expect fewer post-change inconsistencies under MLM.

\textbf{Control (C1).} Applying \textbf{trivial leaf-level mutations} (operations that touch only leaf elements, modifying neither the metamodel structure in 2LM nor any MLM object that types an instantiation relation or parents a level-introducing specialisation) must leave the paired inconsistency-count difference statistically indistinguishable from zero, verified by a two-sided Wilcoxon signed-rank test~\cite{Wilcoxon1945IndividualCB} that fails to reject at $\alpha = 0.05$.

\paragraph{\textbf{H2 --- Modification effort}}
\textbf{Hypothesis.} For semantically equivalent scenarios, restoring consistency after identical evolution mutations incurs, on average, a smaller paradigm-native modification footprint under MLM than under 2LM. We attribute any such reduction to structural unification, not to differences in per-edit cognitive or operational cost, which lie outside our scope.

\textbf{Rationale.} In 2LM, modification effort is dominated by the cascading migration of conforming models across two artefacts~\cite{DBLP:conf/ecoop/HerrmannsdoerferBJ09,DBLP:journals/jss/KhelladiBHG17} (Section~\ref{sec:background}); in MLM as realised by SLICER the change is expressed once on the defining object, its effect mediated by the recorded instantiation and specialisation relations rather than by manual migration. We therefore expect fewer paradigm-native modifications under MLM.

\textbf{Control (C2).} The same trivial leaf-level mutations (C1) must leave the paired element-modification-count difference statistically indistinguishable from zero, verified by an automated diff over the canonical serialisation of each artefact pair before and after restoration.

\paragraph{\textbf{Shared control (C3)}}
\revise{The \textbf{round-trip mapping check} translates a 2LM scenario into its MLM counterpart under the pre-registered mapping protocol (Section~\ref{sec:execution_plan}) and back, and must recover the original 2LM artefact up to a pre-registered notion of structural equivalence. The round-trip establishes that the mapping is information-preserving and invertible, but it does not by itself establish semantic equivalence between the two representations. We therefore pair it with a pre-registered conformance-preservation check: a fixed set of instance-level conformance tests must return identical accept/reject verdicts on the 2LM model and on its MLM counterpart, so that an instance valid under one representation is valid under the other. On a stratified random sample, semantic faithfulness is additionally audited by an independent reviewer blinded to the hypotheses, and any scenario failing either check is excluded. Together these ensure the MLM artefact is a faithful counterpart of the original rather than a paradigm-favouring re-encoding.} Failure on any control invalidates the corresponding test for the affected condition and is reported in the Stage~2 paper as required by the RR protocol.

\section{Variables}\label{sec:variables}
This section operationalises every variable underlying H1, H2, and the controls. Independent and dependent variables are deterministically computable from the artefacts, so that two researchers applying the protocol to the same scenario obtain identical measurements; confounders are listed with the strategy that neutralises or absorbs them. Table~\ref{tab:notation} summarises the notation.

\begin{table}[t]
\caption{Notation used throughout the paper.}
\label{tab:notation}
\centering
\footnotesize
\begin{tabular}{@{}l p{0.66\columnwidth}@{}}
\toprule
\multicolumn{2}{@{}l}{\textit{Variables}}\\
\textsc{Paradigm}    & Modelling paradigm, $\in\{\textsc{2LM},\textsc{MLM}\}$ \\
\textsc{InconsCount} & Post-change inconsistency count (primary outcome, H1) \\
\textsc{ModCount}    & Element modification count (primary outcome, H2) \\
\textsc{Complexity}  & Scenario complexity stratum (small/medium/large) \\
\midrule
\multicolumn{2}{@{}l}{\textit{SLICER relation kinds}}\\
InstN & Standard instantiation \\
InstX & Instantiation with extension \\
SpecX & Specialisation by extension \\
SpecR & Specialisation by refinement \\
SbS   & Subset-by-specification \\
\bottomrule
\end{tabular}
\end{table}

\subsection{Independent Variable}

\subsubsection{\textbf{Modelling Paradigm} (\textsc{Paradigm})}
\textit{Description.} The MDE paradigm under which a co-evolution scenario is represented and evolved.

\noindent\textit{Scale type.} Nominal, two levels.

\noindent\textit{Operationalisation.} 
\textsc{2LM} and \textsc{MLM} scenarios are represented and evolved under EMF and SLICER~\cite{DBLP:journals/dke/SelwaySMJGS17} respectively, with DoME as the MLM reference implementation (see Section~\ref{sec:background} for details).

\subsection{Dependent Variables}
\label{sec:dependent_var}
\subsubsection{\textbf{Post-Change Inconsistency Count} (\textsc{InconsCount}; primary outcome for H1)}
\textit{Description.} The number of distinct consistency violations reported by automated checking on the artefact(s) after applying an evolution mutation, before any restoration is attempted.

\noindent\textit{Scale type.} Ratio (non-negative integer count).

\noindent\textit{Operationalisation.} On each side, \textsc{InconsCount} is the count of violations reported by the paradigm's native validator under a pre-registered, fixed configuration: for \textsc{2LM}, Ecore well-formedness, model-to-metamodel conformance, and OCL invariant violations via the EMF Validation Framework; for \textsc{MLM}, linguistic-metamodel and object-constraint violations via DoME's validator. The exact configurations are pre-registered in the replication package. 

\subsubsection{\textbf{Element Modification Count} (\textsc{ModCount}; primary outcome for H2)}
\textit{Description.} The number of paradigm-native edits needed to return the artefact(s) from the post-mutation state to a consistent one, judged by the same checker configuration used for \textsc{InconsCount}. ``Paradigm-native'' means edits counted at the native granularity of each paradigm's reference tooling (EMF for 2LM, DoME for MLM). Cross-paradigm comparisons of \textsc{ModCount} therefore compare structural edit footprint, not per-edit human labour.

\noindent\textit{Scale type.} Ratio (non-negative integer count).

\noindent\textit{Operationalisation.} \textsc{ModCount} is computed by EMF~Compare (2LM) and DoME's structural diff (MLM) over canonical serialisations before and after restoration, counting four atomic operations (\textsc{add}, \textsc{delete}, \textsc{update}, \textsc{move}) with equal weight over each paradigm's native elements (classes, attributes, references, instances, slots in 2LM; objects, attributes, slots, and the five SLICER relation kinds in MLM). Restoration is deterministic and pre-registered (Section~\ref{sec:execution_plan}), with no human-in-the-loop migration.

\subsection{Confounding Variables}

\subsubsection{\textbf{Scenario Complexity} (\textsc{Complexity})}
\textit{Description.} The intrinsic size and structural richness of a scenario, which may raise both \textsc{InconsCount} and \textsc{ModCount} independently of \textsc{Paradigm}. \textit{Scale type:} ordinal, three strata (\textsc{small}, \textsc{medium}, \textsc{large}). \textit{Control strategy:} stratified sampling. Using pre-registered thresholds on element count and on the depth of the longest level-introducing relation chain (InstN/InstX/SpecX chains in MLM; metamodel-to-model classification depth in 2LM), the corpus (Section~\ref{sec:datasets}) is partitioned into strata, each contributing a balanced number of pairs; the paired Wilcoxon test (Section~\ref{sec:execution_plan}) further absorbs scenario-level variation, since each scenario contributes both a 2LM and an MLM observation.

\subsubsection{\textbf{Mutation Type}}
\textit{Description.} The category of evolution operation applied to a scenario; because different operations exert different synchronisation pressure in 2LM, it could drive the H1/H2 effects asymmetrically if left uncontrolled. \textit{Scale type:} nominal, enumerated by the pre-registered mutation taxonomy (Section~\ref{sec:execution_plan}), derived from Herrmannsdoerfer et al.'s catalogue~\cite{DBLP:conf/sle/HerrmannsdoerferVW10} and adapted to SLICER objects and the five relation kinds (InstN, InstX, SpecX, SpecR, SbS). \textit{Control strategy:} each scenario receives the full set of pre-registered mutations (full-factorial within scenario) wherever mechanically applicable, every paired difference formed within a matched \textit{(scenario, mutation)} cell so that mutation type is identical across paradigms within each pair. \revise{We do not balance across mutation-type categories, and balancing is not required: the primary tests operate on within-pair differences in which mutation type is constant by construction, so any main effect cancels and cannot confound the paradigm comparison. To keep the residual frequency imbalance transparent, we pre-register reporting of the realised count of applicable mutations per category and per \textsc{Complexity} stratum, and report per-mutation effects as a planned secondary breakdown.} The primary tests aggregate across mutation types within scenario pairs.

\subsubsection{\textbf{Mapping Protocol Bias}}
\textit{Description.} Asymmetries inadvertently introduced when translating a 2LM scenario into its SLICER counterpart under the meta-property decomposition of~\cite[Section~6]{DBLP:journals/dke/SelwaySMJGS17}, which could artificially favour MLM. \textit{Control strategy:} \revise{the mapping protocol is fully pre-registered before any scenario is mapped and is applied by an operator blinded 
to the mutation set. Although the operator is aware of the study's directional hypotheses, as is unavoidable given their role in the research team, operator discretion is structurally bounded by the deterministic meta-property decomposition (Section~\ref{sec:execution_plan}), in which the annotation pattern uniquely determines the SLICER relation kind.
To bound the residual latitude in the annotation step itself, a second operator, likewise blinded to the hypotheses, independently annotates a pre-registered random sample of scenarios, and inter-annotator agreement (Cohen's $\kappa$) is reported, with any disagreement resolved by the pre-registered tie-breaking rules. The mapping is additionally validated by the round-trip check (Section~\ref{sec:hypotheses}).}

\subsubsection{\textbf{Tool Maturity}}
\textit{Description.} The greater maturity and wider adoption of EMF (the 2LM reference tooling) over DoME (the SLICER reference tooling) may affect the accuracy of consistency-violation and structural-diff detection. \textit{Control strategy:} validators and diff tools on both sides are configured to a consistent, pre-registered semantic scope (well-formedness, conformance, and invariants; element-level structural diff); to compensate for DoME's relative immaturity, every MLM-side \textsc{InconsCount} and \textsc{ModCount} measurement is subject to a pre-registered manual audit by an independent reviewer blinded to the corresponding 2LM measurement, which may flag a measurement as \textit{instrument failure} (excluding it from the test) but cannot revise its value, with all decisions logged in the replication package. \textit{Residual threat and bias direction:} the audit cannot rule out that DoME under-reports violations relative to EMF, so we bound this by direction: under-reporting biases \textsc{InconsCount} \emph{downward} for MLM, the direction H1 predicts, so a null or 2LM-favouring result on H1 is robust, and a significant MLM-favouring result is reported in Stage~2 with the caveat that part of the reduction may reflect validator coverage; the replication package publishes every violation flagged by either validator for independent scrutiny.

\subsubsection{\textbf{Cross-Paradigm Edit Granularity}}
\textit{Description.} \textsc{InconsCount} and \textsc{ModCount} are counted in each paradigm's native units, and since 2LM and MLM expose different element kinds (coupled metamodel-and-model edits versus single-artefact edits over SLICER's relations), one conceptual change may decompose into a different number of native units in each. \textit{Control strategy:} \textsc{ModCount} captures structural edit footprint at each paradigm's native granularity, not human labour (Section~\ref{sec:dependent_var}), so any paradigm-level asymmetry is read as evidence about structural cost, not person-time; to keep the mapping from amplifying this asymmetry, the mutation taxonomy is anchored at the conceptual-change level (e.g., ``introduce a mandatory attribute on concept $C$'') and instantiated independently against each paradigm's native operations, and the round-trip check (C3) confirms this anchoring introduces no MLM-favouring decomposition. The atomic operations counted by DoME's diff are of comparable fineness to EMF Compare's, so \textsc{ModCount} differences are not expected to be dominated by tool granularity. \textit{Residual threat:} as with Tool Maturity, coarser MLM operations would bias \textsc{ModCount} \emph{downward} for MLM, the direction H2 predicts, so a null or 2LM-favouring result is robust and a significant MLM-favouring result is reported in Stage~2 with the corresponding caveat. 

\section{Datasets}\label{sec:datasets}
This study involves no human participants. Its empirical units are \textit{co-evolution scenarios} drawn from publicly available, peer-reviewed sources in the MDE literature. We deliberately rely on third-party scenario sources rather than constructing scenarios ourselves, so that selection is not influenced by knowledge of subsequent mutations or anticipated paradigm-level outcomes.

\paragraph{\textbf{Source Corpora}}
The corpus is assembled from four public sources of complementary character:
\textit{(i)} large-scale, industrial-character scenarios from the long-running GMF evolution history~\cite{DBLP:conf/sle/HerrmannsdoerferRW09};
\textit{(ii)} canonical small-scale scenarios from Herrmannsdoerfer et al.'s coupled-operator catalogue~\cite{DBLP:conf/sle/HerrmannsdoerferVW10}, which also defines our mutation taxonomy (Mutation Type, Section~\ref{sec:variables});
\textit{(iii)} medium-complexity scenarios with known migration solutions from Hebig et al.'s co-evolution survey~\cite{DBLP:journals/tse/HebigKB17};
\textit{(iv)} small- and medium-scale scenarios from Iovino et al.'s multi-language metamodel-evolution impact study~\cite{DBLP:journals/jot/IovinoPM12}.

\paragraph{\textbf{Selection Criteria}}
Evaluated by the corpus operator \emph{before} any mutation, a scenario is eligible if: (i) its original 2LM artefacts (a metamodel and at least one conforming model) are publicly available or reconstructible; (ii) it specifies an unambiguous before-state; (iii) it is expressible under the SLICER mapping protocol (Section~\ref{sec:execution_plan}) without out-of-scope constructs; and (iv) at least one element can receive at least one mutation from the pre-registered taxonomy. A scenario is excluded if its sources are incomplete or inconsistent, if it relies on EMF tooling extensions with no documented semantics, or if it is a syntactic variant of an accepted scenario (only the more general variant is kept).

\paragraph{\textbf{Target Sample Size and Stratification}}
The target number of paired scenarios is determined by the pre-registered statistical power analysis described in Section~\ref{sec:execution_plan}. Based on the published volumes of the four sources, we expect the number of scenarios meeting our eligibility criteria to exceed the sample size of $N = 55$ required for adequate power. \revise{As an early feasibility gate, we screen the corpus against the eligibility criteria \emph{before} any mutation is applied and record the realised eligible count per source and per \textsc{Complexity} stratum. If this screen projects fewer than $N = 47$ eligible scenarios, we enact the pre-registered source-expansion contingency (secondary sources, in the priority order fixed in the replication package) before proceeding; if expansion still falls short, we report the shortfall and restrict primary inference to the strata that remain adequately powered, documenting the deviation in the Stage~2 paper. The final eligible count is reported in the Stage~2 paper.}

Each eligible scenario is assigned to a \textsc{Complexity} stratum using the pre-registered thresholds of Section~\ref{sec:variables}, and strata are populated to the per-stratum minimum implied by the power analysis, drawing on the secondary sources where a primary-source stratum falls short.

\paragraph{\textbf{Data Availability}}
All artefacts and analysis scripts will be released in a public replication package upon Stage~1 acceptance, per the ICSME open science policy.


\paragraph{\textbf{Threats to Corpus Validity}}
The four primary sources are concentrated in the Eclipse/EMF ecosystem, which may bias the corpus toward 2LM idioms favoured by that toolchain and away from corpora developed under GME, MetaEdit+, or other 2LM frameworks~\cite{zhang2026development}. To mitigate this, the Hebig et al. survey corpus~\cite{DBLP:journals/tse/HebigKB17} contributes scenarios drawn from heterogeneous co-evolution approaches across multiple research groups, and the Iovino et al. study~\cite{DBLP:journals/jot/IovinoPM12} spans multiple modelling languages beyond GMF. A residual threat is that scenarios published in the literature may over-represent ``hard'' co-evolution cases that motivated the original publications, inflating absolute \textsc{InconsCount} and \textsc{ModCount} values. Because the primary tests are paired within scenarios, this absolute inflation does not bias the paradigm-level comparison; we report descriptive statistics on absolute magnitudes alongside the paired tests so that readers can assess external validity directly. \revise{A distinct concern is that scenarios sourced from EMF-oriented communities may embody 2LM design conventions and thus sample the region of the design space where MLM's advantages are least pronounced. This biases the comparison \emph{against} MLM: a demonstrated MLM advantage on such a corpus is a conservative lower bound, while the corpus cannot manufacture an advantage that does not exist. The complementary question, whether MLM's benefits are larger on models designed natively for multi-level representation, falls outside the scope of this confirmatory test and is left to the future work of Section~\ref{sec:future}.}

\section{Execution Plan}\label{sec:execution_plan}
\noindent \textbf{Pipeline.} The execution proceeds in five pre-registered stages, applied uniformly to every eligible scenario:
(1)~\textit{Mapping}: \revise{an operator blinded to the mutation set and to the directional hypotheses} translates the 2LM scenario into its SLICER counterpart under the pre-registered mapping protocol, which applies the meta-property decomposition of Selway et al.~\cite[Section~6]{DBLP:journals/dke/SelwaySMJGS17} so that each construct's annotation pattern determines its SLICER relation kind. The round-trip check (C3) is run on the result, and any failure terminates inclusion.
(2)~\textit{Mutation application}: each mechanically applicable mutation from the pre-registered taxonomy is applied independently to the 2LM and SLICER artefacts, producing one mutated artefact pair per (scenario, mutation).
(3)~\textit{Inconsistency measurement}: \textsc{InconsCount} is recorded immediately after mutation, before any restoration, using the pre-registered EMF Validation Framework configuration (2LM) and DoME validator configuration (MLM).
(4)~\textit{Restoration}: a deterministic, pre-registered restoration procedure resolves violations by applying the minimum-edit closure under a fixed priority of repair operators; restoration is fully scripted and admits no human-in-the-loop decisions.
(5)~\textit{Modification measurement}: \textsc{ModCount} is computed by EMF~Compare (2LM) and DoME's structural diff (MLM) over canonical serialisations of the artefacts before and after restoration. 

\noindent \textbf{Restoration procedure.} 
The restoration script systematically resolves violations in the post-mutation artefact to produce a fully consistent state; the modification distance is reported as \textsc{ModCount} (Section~\ref{sec:dependent_var}).
The script proceeds violation-by-violation in a pre-registered traversal order. For each violation, it selects a repair from a fixed catalogue ranked by a pre-registered priority list. 
The priority ordering encodes two standard maintenance heuristics: prefer non-destructive over destructive repairs, and local over widely propagating ones. The priority list is structurally identical across paradigms but instantiated against the artefact kinds available on each side. The SLICER-side instantiation reuses the semantic conflict resolution catalogue developed in prior work~\cite{DBLP:conf/models/FuGKSS25}.  \revise{The 2LM-side instantiation draws its repairs from the established EMF coupled-evolution operators~\cite{DBLP:conf/sle/HerrmannsdoerferVW10}, the community-standard migration catalogue for that paradigm, so that neither side is restored with an ad hoc operator set. Both catalogues are constrained to the same minimum-edit objective, so \textsc{ModCount} reflects the edit footprint required to regain consistency rather than the relative sophistication of either catalogue. We nonetheless acknowledge a residual asymmetry, that the SLICER catalogue~\cite{DBLP:conf/models/FuGKSS25} was developed by some of the present authors whereas the 2LM operators are third-party, and we bound it by direction: a more effective MLM catalogue can only lower \textsc{ModCount} for MLM, the direction H2 already predicts, so it cannot manufacture a spurious 2LM result nor mask a real one. Should H2 favour MLM, robustness check~(iii) below re-tests it with the authored catalogue removed.} The highest-priority repair that yields a syntactically valid result is applied. Ties are broken lexicographically over repair names. Restoration terminates when no violations remain, or when a pre-registered iteration cap is reached, in which case the scenario is flagged as \textit{unrestorable} and excluded from H2 (but retained for H1, since \textsc{InconsCount} is measured before restoration). The unrestorable rate is reported per paradigm and per stratum as a quality indicator.

\noindent \textbf{Statistical analysis.} H1 and H2 are tested by paired one-sided Wilcoxon signed-rank tests with Holm correction~\cite{Holm1979ASS} across the two primary tests at $\alpha = 0.05$. Effect sizes are reported as the matched-pairs rank-biserial correlation $r_{rb}$ with $95\%$ bootstrap confidence intervals. \revise{We use $r_{rb}$ because it is native to the Wilcoxon signed-rank test: it is computed from the test's signed-rank statistic and therefore quantifies exactly the paired comparison the significance test evaluates, it inherits the test's distribution-free assumptions (unlike a parametric $d_z$, which our skewed, non-negative counts would violate), and it is bounded in $[-1,+1]$ on Cohen's small/medium/large scale, commensurate with the smallest effect of interest $r_{rb} = 0.40$ used below. Its confidence interval is obtained by nonparametric bootstrap over scenario pairs.} Outcome-neutral controls C1 and C2 (Section~\ref{sec:hypotheses}) are evaluated as two-sided Wilcoxon tests under the same $\alpha$, and primary inference is conditional on failure to reject in C1, C2, and the structural pass of C3. Scenario--mutation pairs flagged as \textit{instrument failure} by the manual audit are excluded from the corresponding hypothesis test; the instrument-failure rate is reported per paradigm and per stratum.

\noindent \textbf{Power analysis.} Sample size is determined a priori for the more conservative of the two primary tests. Targeting power $1 - \beta = 0.80$ at the Holm-corrected level $\alpha' = 0.025$ with a one-sided paired Wilcoxon test, and a pre-registered smallest effect of interest of $r_{rb} = 0.40$ (a medium effect by Cohen's conventions~\cite{cohen1988}, in line with SE experiment benchmarks~\cite{DBLP:journals/infsof/KampenesDHS07}), the required minimum sample is $N = 47$ paired scenarios. To absorb potential attrition from manual-audit instrument-failure flags, we pre-register a target sample size of $N = 55$ ($\approx 17\%$ buffer); should the post-audit effective sample fall below $N = 47$, the deviation will be reported in the Stage~2 paper. 

\noindent \textbf{Robustness checks.} Three pre-registered sensitivity analyses accompany the primary inference. (i) An alternative restoration priority list, swapping the top two repair operators, is run on the full corpus to verify that H2 conclusions are not driven by the specific priority pre-registration. (ii) A leave-one-source-out re-analysis re-computes the primary tests with each of the four corpus sources omitted in turn, exposing any single-source dominance.  \revise{(iii) H2 is re-computed with the SLICER-side restoration restricted to a paradigm-generic, minimum-edit repair set that excludes the semantic conflict-resolution catalogue of~\cite{DBLP:conf/models/FuGKSS25} and mirrors the generic operators used on the 2LM side, confirming that any MLM advantage does not depend on the authored catalogue.}
Disagreement between the primary analysis and any robustness check is reported in the Stage~2 paper without retro-fitting hypotheses.

\section{Work in Progress and Future Work}\label{sec:future}
\revise{This manuscript is a Stage~1 Registered Report: it fixes the design, hypotheses, and analysis plan, while the confirmatory results are reported in the Stage~2 paper. The protocol is therefore presented without outcome data by design. Work to date has fixed the variable operationalisation (Section~\ref{sec:variables}) and the corpus eligibility criteria (Section~\ref{sec:datasets}); the mapping protocol, mutation taxonomy, restoration procedure, and validator configurations are being finalised and frozen.}

Following Stage~1 In-Principle Acceptance, Stage~2 execution is scheduled over approximately twelve months, leaving a buffer before the EMSE submission deadline of September~11, 2027:
\begin{itemize}
  \item \textit{Months 1--2.} Freeze the replication package: mapping protocol, mutation taxonomy over the SLICER relation kinds, restoration procedure, and validator configurations.
  \item \textit{Months 3--5.} Screen the corpus against the eligibility criteria and run an early feasibility gate: if the projected eligible count falls below $N = 47$, enact the pre-registered source-expansion contingency before proceeding. Build SLICER counterparts under the blinded mapping protocol, including round-trip verification (C3).
  \item \textit{Months 6--8.} Apply the pre-registered mutations and measure \textsc{InconsCount} and \textsc{ModCount} automatically.
  \item \textit{Months 9--10.} Evaluate the outcome-neutral controls (C1, C2), then run the Holm-corrected tests of H1 and H2 and the planned per-stratum breakdown.
  \item \textit{Months 11--12.} Manuscript preparation, internal review, and submission.
\end{itemize}
Any deviation from this plan that materially affects the protocol will be documented in the Stage~2 paper.

Beyond Stage~2, the design generalises in two directions. \revise{First, the mapping protocol can be re-targeted to other multi-level frameworks, testing whether the measured maintenance effects are specific to SLICER or hold across MLM realisations. Second, because \textsc{ModCount} measures structural edit footprint rather than engineering effort, a reduction in edits is a necessary but not sufficient indicator of reduced maintenance cost. Establishing how structural footprint maps to developer time requires controlled human-subject studies, which, together with richer edit-cost models and a broader set of evolution operators, we leave to future work. Complementary approaches that leverage LLMs to support co-evolution~\cite{zhang2026leveraging} suggest further directions for automating the restoration step.}



\newpage
\bibliographystyle{IEEEtran}
\bibliography{icsme26}

@article{DBLP:journals/computer/Schmidt06,
  author       = {Douglas C. Schmidt},
  title        = {Guest Editor's Introduction: Model-Driven Engineering},
  journal      = {Computer},
  volume       = {39},
  number       = {2},
  pages        = {25--31},
  year         = {2006},
  _url          = {https://doi.org/10.1109/MC.2006.58},
  doi          = {10.1109/MC.2006.58},
  timestamp    = {Wed, 12 Aug 2020 10:30:10 +0200},
  biburl       = {https://dblp.org/rec/journals/computer/Schmidt06.bib},
  bibsource    = {dblp computer science bibliography, https://dblp.org}
}

@book{DBLP:series/synthesis/2017Brambilla,
  author       = {Marco Brambilla and
                  Jordi Cabot and
                  Manuel Wimmer},
  title        = {Model-Driven Software Engineering in Practice, Second Edition},
  series       = {Synthesis Lectures on Software Engineering},
  publisher    = {Morgan {\&} Claypool Publishers},
  year         = {2017},
  _url          = {https://doi.org/10.2200/S00751ED2V01Y201701SWE004},
  doi          = {10.2200/S00751ED2V01Y201701SWE004},
  isbn         = {978-3-031-01421-5},
  timestamp    = {Sun, 19 Jan 2025 15:07:15 +0100},
  biburl       = {https://dblp.org/rec/series/synthesis/2017Brambilla.bib},
  bibsource    = {dblp computer science bibliography, https://dblp.org}
}

@inproceedings{DBLP:conf/edoc/CicchettiREP08,
  author       = {Antonio Cicchetti and
                  Davide Di Ruscio and
                  Romina Eramo and
                  Alfonso Pierantonio},
  title        = {Automating Co-evolution in Model-Driven Engineering},
  booktitle    = {{ECOC} 2008},
  pages        = {222--231},
  publisher    = {{IEEE} Computer Society},
  year         = {2008},
  _url          = {https://doi.org/10.1109/EDOC.2008.44},
  doi          = {10.1109/EDOC.2008.44},
  timestamp    = {Mon, 05 Feb 2024 20:33:49 +0100},
  biburl       = {https://dblp.org/rec/conf/edoc/CicchettiREP08.bib},
  bibsource    = {dblp computer science bibliography, https://dblp.org}
}

@inproceedings{DBLP:conf/ecoop/HerrmannsdoerferBJ09,
  author       = {Markus Herrmannsdoerfer and
                  Sebastian Benz and
                  Elmar J{\"{u}}rgens},
  _editor       = {Sophia Drossopoulou},
  title        = {{COPE} - Automating Coupled Evolution of Metamodels and Models},
  booktitle    = {Proc. of {ECOOP} 2009},
  series       = {LNCS},
  pages        = {52--76},
  publisher    = {Springer},
  year         = {2009},
  _url          = {https://doi.org/10.1007/978-3-642-03013-0\_4},
  doi          = {10.1007/978-3-642-03013-0\_4},
  timestamp    = {Tue, 14 May 2019 10:00:54 +0200},
  biburl       = {https://dblp.org/rec/conf/ecoop/HerrmannsdoerferBJ09.bib},
  bibsource    = {dblp computer science bibliography, https://dblp.org}
}

@inproceedings{DBLP:conf/sle/HerrmannsdoerferVW10,
  author       = {Markus Herrmannsdoerfer and
                  Sander Vermolen and
                  Guido Wachsmuth},
  _editor       = {Brian A. Malloy and
                  Steffen Staab and
                  Mark van den Brand},
  title        = {An Extensive Catalog of Operators for the Coupled Evolution of Metamodels
                  and Models},
  booktitle    = {{SLE} 2010},
  series       = {LNCS},
  pages        = {163--182},
  publisher    = {Springer},
  year         = {2010},
  _url          = {https://doi.org/10.1007/978-3-642-19440-5\_10},
  doi          = {10.1007/978-3-642-19440-5\_10},
  timestamp    = {Fri, 27 Dec 2019 21:19:31 +0100},
  biburl       = {https://dblp.org/rec/conf/sle/HerrmannsdoerferVW10.bib},
  bibsource    = {dblp computer science bibliography, https://dblp.org}
}

@article{DBLP:journals/jss/KhelladiBHG17,
  author       = {Djamel Eddine Khelladi and
                  Reda Bendraou and
                  Regina Hebig and
                  Marie{-}Pierre Gervais},
  title        = {A semi-automatic maintenance and co-evolution of {OCL} constraints
                  with (meta)model evolution},
  journal      = {J. Syst. Softw.},
  volume       = {134},
  pages        = {242--260},
  year         = {2017},
  _url          = {https://doi.org/10.1016/j.jss.2017.09.010},
  doi          = {10.1016/J.JSS.2017.09.010},
  timestamp    = {Sun, 19 Jan 2025 14:04:10 +0100},
  biburl       = {https://dblp.org/rec/journals/jss/KhelladiBHG17.bib},
  bibsource    = {dblp computer science bibliography, https://dblp.org}
}

@article{DBLP:journals/tosem/LaraGC14,
  author       = {Juan de Lara and
                  Esther Guerra and
                  Jes{\'{u}}s S{\'{a}}nchez Cuadrado},
  title        = {When and How to Use Multilevel Modelling},
  journal      = {{ACM} Trans. Softw. Eng. Methodol.},
  volume       = {24},
  number       = {2},
  pages        = {12:1--12:46},
  year         = {2014},
  _url          = {https://doi.org/10.1145/2685615},
  doi          = {10.1145/2685615},
  timestamp    = {Tue, 06 Nov 2018 12:51:20 +0100},
  biburl       = {https://dblp.org/rec/journals/tosem/LaraGC14.bib},
  bibsource    = {dblp computer science bibliography, https://dblp.org}
}

@article{DBLP:journals/bise/Frank14,
  author       = {Ulrich Frank},
  title        = {Multilevel Modeling - Toward a New Paradigm of Conceptual Modeling
                  and Information Systems Design},
  journal      = {Bus. Inf. Syst. Eng.},
  volume       = {6},
  number       = {6},
  pages        = {319--337},
  year         = {2014},
  _url          = {https://doi.org/10.1007/s12599-014-0350-4},
  doi          = {10.1007/S12599-014-0350-4},
  timestamp    = {Fri, 06 Mar 2020 21:58:39 +0100},
  biburl       = {https://dblp.org/rec/journals/bise/Frank14.bib},
  bibsource    = {dblp computer science bibliography, https://dblp.org}
}

@article{DBLP:journals/dke/SelwaySMJGS17,
  author       = {Matt Selway and
                  Markus Stumptner and
                  Wolfgang Mayer and
                  Andreas Jordan and
                  Georg Grossmann and
                  Michael Schrefl},
  title        = {A conceptual framework for large-scale ecosystem interoperability
                  and industrial product lifecycles},
  journal      = {Data Knowl. Eng.},
  volume       = {109},
  pages        = {85--111},
  year         = {2017},
  _url          = {https://doi.org/10.1016/j.datak.2017.03.006},
  doi          = {10.1016/J.DATAK.2017.03.006},
  timestamp    = {Mon, 28 Aug 2023 21:41:49 +0200},
  biburl       = {https://dblp.org/rec/journals/dke/SelwaySMJGS17.bib},
  bibsource    = {dblp computer science bibliography, https://dblp.org}
}

@inproceedings{DBLP:conf/sle/HerrmannsdoerferRW09,
  author       = {Markus Herrmannsdoerfer and
                  Daniel Ratiu and
                  Guido Wachsmuth},
  _editor       = {Mark van den Brand and
                  Dragan Gasevic and
                  Jeff Gray},
  title        = {Language Evolution in Practice: The History of {GMF}},
  booktitle    = {{SLE} 2009},
  series       = {LNCS},
  pages        = {3--22},
  publisher    = {Springer},
  year         = {2009},
  _url          = {https://doi.org/10.1007/978-3-642-12107-4\_3},
  doi          = {10.1007/978-3-642-12107-4\_3},
  timestamp    = {Mon, 21 Jun 2021 12:26:17 +0200},
  biburl       = {https://dblp.org/rec/conf/sle/HerrmannsdoerferRW09.bib},
  bibsource    = {dblp computer science bibliography, https://dblp.org}
}

@article{DBLP:journals/tse/HebigKB17,
  author       = {Regina Hebig and
                  Djamel Eddine Khelladi and
                  Reda Bendraou},
  title        = {Approaches to Co-Evolution of Metamodels and Models: {A} Survey},
  journal      = {{IEEE} Trans. Software Eng.},
  volume       = {43},
  number       = {5},
  pages        = {396--414},
  year         = {2017},
  _url          = {https://doi.org/10.1109/TSE.2016.2610424},
  doi          = {10.1109/TSE.2016.2610424},
  timestamp    = {Sun, 06 Oct 2024 21:41:51 +0200},
  biburl       = {https://dblp.org/rec/journals/tse/HebigKB17.bib},
  bibsource    = {dblp computer science bibliography, https://dblp.org}
}

@article{DBLP:journals/jot/IovinoPM12,
  author       = {Ludovico Iovino and
                  Alfonso Pierantonio and
                  Ivano Malavolta},
  title        = {On the Impact Significance of Metamodel Evolution in {MDE}},
  journal      = {J. Object Technol.},
  volume       = {11},
  number       = {3},
  pages        = {3: 1--33},
  year         = {2012},
  _url          = {https://doi.org/10.5381/jot.2012.11.3.a3},
  doi          = {10.5381/JOT.2012.11.3.A3},
  timestamp    = {Wed, 17 Feb 2021 08:57:54 +0100},
  biburl       = {https://dblp.org/rec/journals/jot/IovinoPM12.bib},
  bibsource    = {dblp computer science bibliography, https://dblp.org}
}

@ARTICLE{1456074,
  author={Lehman, M.M.},
  journal={Proceedings of the IEEE}, 
  title={Programs, life cycles, and laws of software evolution}, 
  year={1980},
  volume={68},
  number={9},
  pages={1060-1076},
  doi={10.1109/PROC.1980.11805}}

@article{DBLP:journals/infsof/Sjoberg93,
  author       = {Dag I. K. Sj{\o}berg},
  title        = {Quantifying schema evolution},
  journal      = {Inf. Softw. Technol.},
  volume       = {35},
  number       = {1},
  pages        = {35--44},
  year         = {1993},
  _url          = {https://doi.org/10.1016/0950-5849(93)90027-Z},
  doi          = {10.1016/0950-5849(93)90027-Z},
  timestamp    = {Mon, 18 Mar 2024 16:37:18 +0100},
  biburl       = {https://dblp.org/rec/journals/infsof/Sjoberg93.bib},
  bibsource    = {dblp computer science bibliography, https://dblp.org}
}

@inproceedings{DBLP:conf/sigsoft/RobbesLR12,
  author       = {Romain Robbes and
                  Mircea Lungu and
                  David R{\"{o}}thlisberger},
  _editor       = {Will Tracz and
                  Martin P. Robillard and
                  Tevfik Bultan},
  title        = {How do developers react to {API} deprecation?: the case of a smalltalk
                  ecosystem},
  booktitle    = {{SIGSOFT} 2012},
  pages        = {56},
  publisher    = {{ACM}},
  year         = {2012},
  _url          = {https://doi.org/10.1145/2393596.2393662},
  doi          = {10.1145/2393596.2393662},
  timestamp    = {Tue, 01 Feb 2022 10:45:16 +0100},
  biburl       = {https://dblp.org/rec/conf/sigsoft/RobbesLR12.bib},
  bibsource    = {dblp computer science bibliography, https://dblp.org}
}

@article{DBLP:journals/smr/DigJ06,
  author       = {Danny Dig and
                  Ralph E. Johnson},
  title        = {How do APIs evolve? {A} story of refactoring},
  journal      = {J. Softw. Maintenance Res. Pract.},
  volume       = {18},
  number       = {2},
  pages        = {83--107},
  year         = {2006},
  _url          = {https://doi.org/10.1002/smr.328},
  doi          = {10.1002/SMR.328},
  timestamp    = {Wed, 20 May 2020 21:25:28 +0200},
  biburl       = {https://dblp.org/rec/journals/smr/DigJ06.bib},
  bibsource    = {dblp computer science bibliography, https://dblp.org}
}

@article{Wilcoxon1945IndividualCB,
  title={Individual Comparisons by Ranking Methods},
  author={Frank. Wilcoxon},
  journal={Biometrics},
  year={1945},
  volume={1},
  pages={196-202},
  url={https://api.semanticscholar.org/CorpusID:53662922}
}

@article{Holm1979ASS,
  title={A Simple Sequentially Rejective Multiple Test Procedure},
  author={Sture Holm},
  journal={Scandinavian Journal of Statistics},
  year={1979},
  volume={6},
  number={2},
  pages={65-70},
  url={http://www.jstor.org/stable/4615733}
}

@inproceedings{DBLP:conf/models/FuGKSS25,
  author       = {Yuhong Fu and
                  Georg Grossmann and
                  Karamjit Kaur and
                  Matt Selway and
                  Markus Stumptner},
  title        = {Conflict Management for Multi-Level Models in Collaborative Modelling
                  Environments},
  booktitle    = {{MODELS} 2025 Companion},
  pages        = {502--511},
  publisher    = {{IEEE}},
  year         = {2025},
  _url          = {https://doi.org/10.1109/MODELS-C68889.2025.00073},
  doi          = {10.1109/MODELS-C68889.2025.00073},
  timestamp    = {Fri, 30 Jan 2026 11:09:34 +0100},
  biburl       = {https://dblp.org/rec/conf/models/FuGKSS25.bib},
  bibsource    = {dblp computer science bibliography, https://dblp.org}
}

@book{cohen1988,
  author    = {Jacob Cohen},
  title     = {Statistical Power Analysis for the Behavioral Sciences},
  edition   = {2nd},
  publisher = {Lawrence Erlbaum Associates},
  address   = {Hillsdale, NJ},
  year      = {1988}
}

@article{DBLP:journals/infsof/KampenesDHS07,
  author       = {Vigdis By Kampenes and
                  Tore Dyb{\aa} and
                  Jo Erskine Hannay and
                  Dag I. K. Sj{\o}berg},
  title        = {A systematic review of effect size in software engineering experiments},
  journal      = {Inf. Softw. Technol.},
  volume       = {49},
  number       = {11-12},
  pages        = {1073--1086},
  year         = {2007},
  _url          = {https://doi.org/10.1016/j.infsof.2007.02.015},
  doi          = {10.1016/J.INFSOF.2007.02.015},
  timestamp    = {Thu, 20 Feb 2020 13:20:48 +0100},
  biburl       = {https://dblp.org/rec/journals/infsof/KampenesDHS07.bib},
  bibsource    = {dblp computer science bibliography, https://dblp.org}
}

@inproceedings{DBLP:conf/modellierung/AtkinsonG16,
  author       = {Colin Atkinson and
                  Ralph Gerbig},
  _editor       = {Stefanie Betz and
                  Ulrich Reimer},
  title        = {Flexible Deep Modeling with Melanee},
  booktitle    = {Modellierung 2016, 2.-4. M{\"{a}}rz 2016, Karlsruhe - Workshopband},
  series       = {{LNI}},
  pages        = {117--122},
  publisher    = {{GI}},
  year         = {2016},
  _url          = {https://dl.gi.de/handle/20.500.12116/843},
  timestamp    = {Tue, 04 Jul 2023 17:44:31 +0200},
  biburl       = {https://dblp.org/rec/conf/modellierung/AtkinsonG16.bib},
  bibsource    = {dblp computer science bibliography, https://dblp.org}
}

@inproceedings{DBLP:conf/models/MaciasRS16,
  author       = {Fernando Mac{\'{\i}}as and
                  Adrian Rutle and
                  Volker Stolz},
  _editor       = {Colin Atkinson and
                  Georg Grossmann and
                  Tony Clark},
  title        = {MultEcore: Combining the Best of Fixed-Level and Multilevel Metamodelling},
  booktitle    = {{MODELS} 2016 Companion (Proc. of MULTI Workshop)},
  series       = {{CEUR} Workshop Proceedings},
  pages        = {66--75},
  publisher    = {CEUR-WS.org},
  year         = {2016},
  _url          = {https://ceur-ws.org/Vol-1722/p6.pdf},
  timestamp    = {Fri, 10 Mar 2023 16:22:20 +0100},
  biburl       = {https://dblp.org/rec/conf/models/MaciasRS16.bib},
  bibsource    = {dblp computer science bibliography, https://dblp.org}
}

@inproceedings{DBLP:conf/er/JeusfeldN16,
  author       = {Manfred A. Jeusfeld and
                  Bernd Neumayr},
  _editor       = {Isabelle Comyn{-}Wattiau and
                  Katsumi Tanaka and
                  Il{-}Yeol Song and
                  Shuichiro Yamamoto and
                  Motoshi Saeki},
  title        = {DeepTelos: Multi-level Modeling with Most General Instances},
  booktitle    = {Proc. of {ER} 2016},
  series       = {LNCS},
  pages        = {198--211},
  year         = {2016},
  _url          = {https://doi.org/10.1007/978-3-319-46397-1\_15},
  doi          = {10.1007/978-3-319-46397-1\_15},
  timestamp    = {Thu, 23 Jun 2022 19:56:58 +0200},
  biburl       = {https://dblp.org/rec/conf/er/JeusfeldN16.bib},
  bibsource    = {dblp computer science bibliography, https://dblp.org}
}

@article{zhang2026leveraging,
  title={Leveraging LLMs to support co-evolution between definitions and instances of textual DSLs: A Systematic Evaluation},
  author={Zhang, Weixing and Jiang, Bowen and Fu, Yuhong and Koziolek, Anne and Hebig, Regina and Str{\"u}ber, Daniel},
  journal={arXiv preprint arXiv:2602.11904},
  year={2026}
}

@article{zhang2026development,
  title={Development and evolution of Xtext-based DSLs on GitHub: an empirical investigation},
  author={Zhang, Weixing and Str{\"u}ber, Daniel and Hebig, Regina},
  journal={Empirical Software Engineering},
  volume={31},
  number={3},
  pages={48},
  year={2026},
  publisher={Springer}
}

@article{zhang2024supporting,
  title={Supporting meta-model-based language evolution and rapid prototyping with automated grammar transformation},
  author={Zhang, Weixing and Holtmann, J{\"o}rg and Str{\"u}ber, Daniel and Hebig, Regina and Stegh{\"o}fer, Jan-Philipp},
  journal={Journal of Systems and Software},
  volume={214},
  pages={112069},
  year={2024},
  publisher={Elsevier}
}

@inproceedings{fu2024modelling,
  author       = {Yuhong Fu and
                  Matt Selway and
                  Georg Grossmann and
                  Karamjit Kaur and
                  Markus Stumptner},
  _editor       = {Manuel Wimmer and
                  Alexander Egyed and
                  Beno{\^{\i}}t Combemale and
                  Marsha Chechik},
  title        = {Modelling a Warehouse with {SLICER:} {A} Contribution to the {MULTI}
                  Warehouse Challenge},
  booktitle    = {MODELS 2024 Companion (Proc. MULTI Workshop)},
  pages        = {828--837},
  publisher    = {{ACM}},
  year         = {2024},
  _url          = {https://doi.org/10.1145/3652620.3688215},
  doi          = {10.1145/3652620.3688215},
  timestamp    = {Wed, 06 Nov 2024 22:17:23 +0100},
  biburl       = {https://dblp.org/rec/conf/models/FuSGKS24.bib},
  bibsource    = {dblp computer science bibliography, https://dblp.org}
}

@inproceedings{DBLP:conf/in4pl/FuGKSS22,
  author       = {Yuhong Fu and
                  Georg Grossmann and
                  Karamjit Kaur and
                  Matt Selway and
                  Markus Stumptner},
  _editor       = {Herv{\'{e}} Panetto and
                  Georg Weichhart and
                  Alexander V. Smirnov and
                  Kurosh Madani},
  title        = {Multi-level Risk Modelling for Interoperability of Risk Information},
  booktitle    = {Proc. of IN4PL 2022},
  pages        = {242--249},
  publisher    = {{SCITEPRESS}},
  year         = {2022},
  _url          = {https://doi.org/10.5220/0011562300003329},
  doi          = {10.5220/0011562300003329},
  timestamp    = {Sat, 30 Sep 2023 09:48:45 +0200},
  biburl       = {https://dblp.org/rec/conf/in4pl/FuGKSS22.bib},
  bibsource    = {dblp computer science bibliography, https://dblp.org}
}

@inproceedings{DBLP:conf/models/FuGKSS23,
  author       = {Yuhong Fu and
                  Georg Grossmann and
                  Karamjit Kaur and
                  Matt Selway and
                  Markus Stumptner},
  title        = {Towards the Integration of Multi-Level and Multi-View Modelling for
                  Interoperability},
  booktitle    = {MODELS 2023 Companion (Proc. MULTI Workshop)},
  pages        = {679--688},
  publisher    = {{IEEE}},
  year         = {2023},
  _url          = {https://doi.org/10.1109/MODELS-C59198.2023.00109},
  doi          = {10.1109/MODELS-C59198.2023.00109},
  timestamp    = {Tue, 07 May 2024 20:10:44 +0200},
  biburl       = {https://dblp.org/rec/conf/models/FuGKSS23.bib},
  bibsource    = {dblp computer science bibliography, https://dblp.org}
}

\end{document}